\documentclass[aps,prl,twocolumn,groupedaddress,10pt]{revtex4-1}
\usepackage{mathrsfs}
\usepackage{amsmath}
\usepackage{amsfonts}
\usepackage{amssymb}
\usepackage{amsthm}
\usepackage{graphicx}
\usepackage{color}
\usepackage{graphics}
\usepackage{epsfig}
\usepackage{ulem}
\usepackage{subfigure}
\usepackage{hyperref}
\providecommand{\U}[1]{\protect\rule{.1in}{.1in}}

\begin{document}
	\title{Current-Driven Dynamics of Magnetic Hopfions}
	\author{X. S. Wang}
	\author{A. Qaiumzadeh}
	\author{A. Brataas}
	
	\affiliation{Center for Quantum Spintronics, Department of Physics,
		Norwegian University of Science and Technology, NO-7491 Trondheim, Norway}

	\begin{abstract}
		Topological magnetic textures have attracted considerable interest since they exhibit new properties and might be useful in information technology. Magnetic hopfions are three-dimensional (3D) spatial variations in the magnetization with a non-trivial Hopf index.
		We find that in ferromagnetic materials, two types of hopfions, Bloch-type and N\'{e}el-type hopfions, can be excited as metastable states in the presence of bulk and interfacial Dzyaloshinskii-Moriya interactions, respectively. We further investigate how hopfions can be driven by currents via spin-transfer torques (STTs) and spin-Hall torques (SHTs). Distinct from 2D ferromagnetic skyrmions, hopfions have a vanishing gyrovector. Consequently, there are no undesirable Hall effects.
		N\'{e}el-type hopfions move along the current direction via both STT and SHE, while Bloch-type hopfions move either transverse to the current direction via SHT or parallel to the current direction via STT.
		Our findings open the door to utilizing hopfions as information carriers.
	\end{abstract}
	
	\maketitle
	
	Topological solitons are of fundamental interest in nonlinear field theories. Additionally, their
	magnetic realizations are promising candidates as information carriers
	in the next generation of data storage and processing devices \cite{soliton1,book}.
	Low-dimensional topological soliton-like textures in ferromagnetic (FM) and antiferromagnetic (AFM)
	materials, such as 1D magnetic domain walls \cite{DW1,DW2,DW3,DW4}, 2D magnetic vortices \cite{VT1,VT2},
	and 2D magnetic skyrmions \cite{SK1,SK2,SK3,SK4,SK5,SKAQ1,SKAQ2,SK6,SK7}, have been extensively studied in recent years.
	
	The existence of 3D topological solitons with string-like properties
	has been proposed by Ludvig D. Faddeev \cite{Faddeev} as a limit of the Skyrme model \cite{Skyrme}.
	These 3D topological solitons are known as Faddeev-Hopf knots \cite{FHK} or hopfions,
	which are classified by a topological charge called the Hopf index \cite{Hp1}.
	Hopfions have been discussed in many physical systems,
	such as gauge theories \cite{Faddeev,Hp2}, cosmic strings \cite{Hp3}, ferromagnets \cite{VR}
	(as a special case of dynamical vortex rings), low-temperature bosonic systems \cite{Egor1,Egor2,Hp4},
	fluids \cite{Hp5}, and liquid crystals \cite{Hp6,Hp7,news}.
	Recently, stable magnetic hopfions were numerically predicted in
	finite-size noncentrosymmetric FM systems with Dzyaloshinskii-Moriya interaction (DMI) \cite{DMI1,DMI2}
	and interfacial perpendicular magnetic anisotropy (PMA)
	\cite{magHp1,magHp2,magHp3} or higher-order exchange interaction \cite{Bluegel2019}.
	However, 3D topological solitons such as hopfions in magnetic systems are still underexplored
	compared to well-studied 1D and 2D solitons.
	
	In this Letter, we show that, in addition to interfacial PMA,
	a bulk PMA assists in stabilizing a localized hopfion that can exist in nanostrips,
	in contrast to the boundary-confined hopfions in nanodisks proposed in previous
	studies \cite{magHp1,magHp2}. In addition to the Bloch-type hopfions studied previously
	\cite{magHp1,magHp2,magHp3}, which can be stable in the presence of bulk DMI
	\cite{DMI1,DMI2,DMI3}, we identify another type of hopfion, N\'{e}el-type hopfions,
	which can be stable in the presence of interfacial DMI \cite{AQDMI}.
	We also introduce an ansatz that can accurately describe the hopfion profile.
	We then study the current-driven dynamics of ferromagnetic hopfions in
	nanostrips. Although the hopfions are topologically nontrivial, their gyrovectors
	vanish. This is in contrast to magnetic skyrmions, whose nontrivial topology induces an
	unwanted ``skyrmion Hall effect" \cite{Hall1,Hall2,Hall3} and hinders the device
	applications \cite{Yoo2017,Device1,Device2}. As a result, hopfions move
	along the current via spin-transfer torques (STTs) \cite{ZhangLi}. Spin Hall torques
	(SHT) \cite{SHT} also cause N\'{e}el-type hopfions to move along the current, while
	Bloch-type hopfions move transverse to the current. Hopfions may be superior to skyrmions
	as information carriers in racetrack memories since their current-induced motion is more straightforward.
	
	We consider a magnetic film of thickness $d$ with interfacial PMA
	at the top and bottom surfaces as well as bulk PMA in the bulk. The zero-temperature micromagnetic
	free energy of the system reads
	\begin{multline}\label{energy}
		\mathcal{F}=\int_V A_{ex} \bigg[\left|\nabla \mathbf{m}\right|^2+\mathscr{D}\left(\mathbf{m},\frac{\partial \mathbf{m}}{\partial x_i}\right)
		+K_b(1 - m_z^2) \\+
		BM_s(1-m_z)\bigg]\mathrm{d} V + \int_{z=\pm d/2} K_s(1-m_z^2) \mathrm{d}S + E_d,
	\end{multline}
	where $A_{ex}$ is the exchange constant; $\mathscr{D}$ is the DMI energy density functional, which
	depends on the symmetry of the system; $K_b$ and $K_s$ are the bulk PMA and the interfacial PMA, respectively;
	$B$ is a perpendicular magnetic field; $M_s$ is the saturation magnetization; and $E_d$ is the demagnetization energy. In bulk noncentrosymmetric materials such as FeGe and MnSi,
	the DMI is bulk-like $\mathscr{D}=D_b \mathbf{m}\cdot(\nabla\times\mathbf{m})$, where $D_b$ is the
	bulk DMI strength in units of J/m$^2$ \cite{SK2}. In inversion-symmetry-broken films such as
	$\text{Pt/Co/AlO}_\text{x}$, the DMI is interfacial-like $\mathscr{D}=D_i\left[(\hat{\mathbf{z}}\cdot
	\mathbf{m})\nabla\cdot\mathbf{m}-(\mathbf{m}\cdot\nabla)(\hat{\mathbf{z}}\cdot\mathbf{m})\right]$,
	where $\hat{\mathbf{z}}$ is the direction normal to the film and $D_i$ is the interfacial DMI
	strength in units of J/m$^2$ \cite{SK1,AQDMI}.
	Because the hopfions are non-isomorphic maps from $\mathbb{R}^3\cup \{\infty\}$ to
	$\mathbb{S}^2$, the topological invariant of hopfions, known as the Hopf index $H$,
	differs from the skyrmion number. This index is defined as
	\begin{equation}\label{hpindex}
	H=\frac{1}{(4\pi)^2}\int_V\mathbf{F}\cdot\mathbf{A}\mathrm{d}V,
	\end{equation}
	where $F_i=\varepsilon_{ijk}\mathbf{m}\cdot\left(\partial_j \mathbf{m}\times\partial_k \mathbf{m}\right)/2$, in which $i,j,k=\{x,y,z\}$ and
	$\varepsilon$ is the Levi-Civita tensor, and $\mathbf{A}$
	is a vector potential, which satisfies $\nabla\times \mathbf{A}=\mathbf{F}$ \cite{HPnumber}. The components of $\mathbf{F}$ are
	solid angle densities in different coordinate planes. $\mathbf{F}$ can be understood as the
	gyrovector density \cite{Thiele}, emergent magnetic field \cite{SZhang2009}, or
	topological charge \cite{SK2}.
	
	\begin{figure}[!t]
		\begin{center}
			\includegraphics[width=8.5cm]{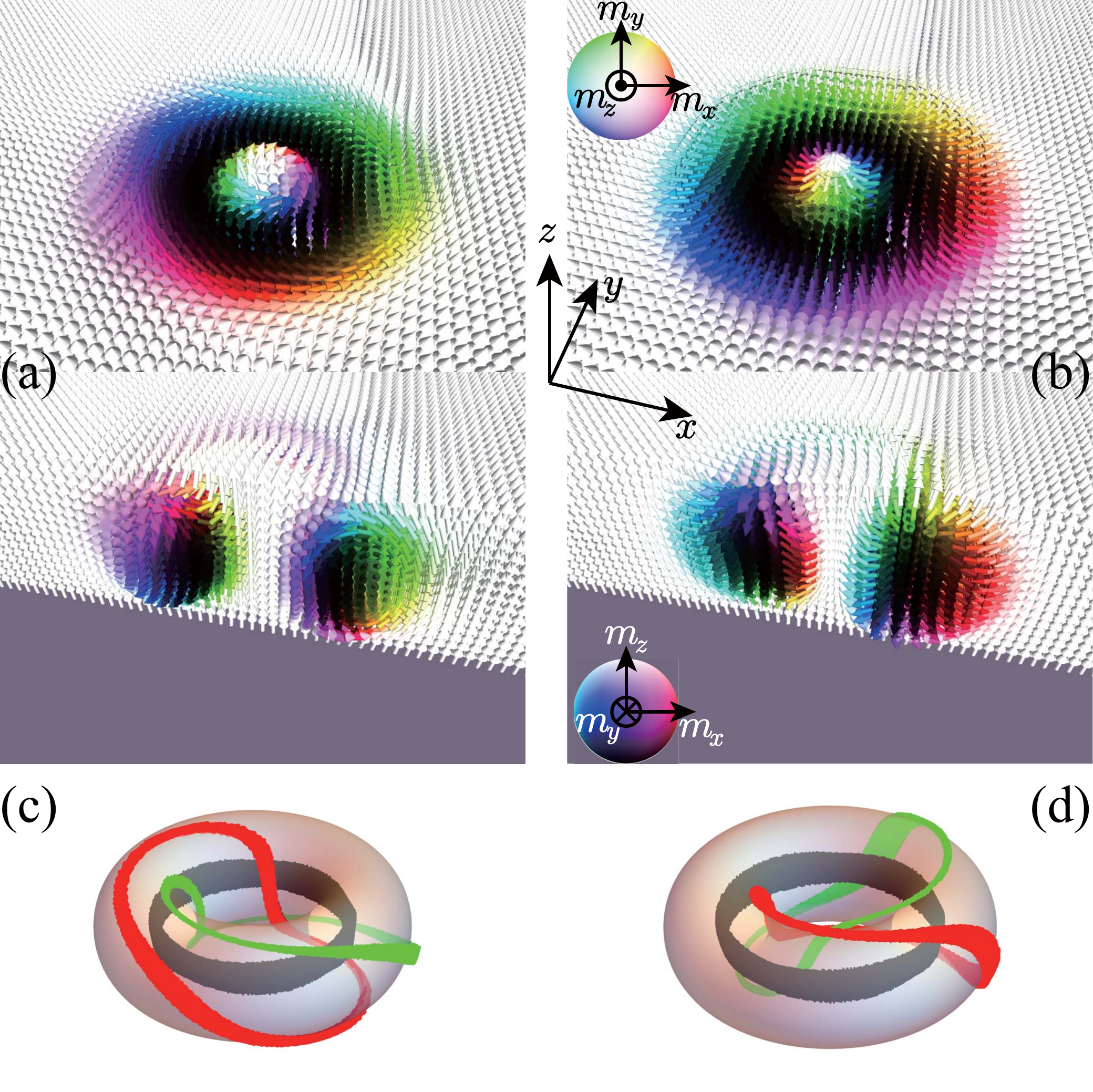}
		\end{center}
		\caption{
			(a)(b) Midplane cross-sections in the $xy$-plane (upper panel) and the $xz$-plane (lower panel)
			of (a) a Bloch-type hopfion and (b) a N\'{e}el-type hopfion.
			(c)(d) The preimages of $\mathbf{m}=(0,0,-1)$, $(1,0,0)$ and $(0,1,0)$
			for (c) a Bloch-type hopfion and (d) a N\'{e}el-type hopfion. The tori
			are the isosurfaces of $m_z=0$.
			The colors of the arrows in (a)(b) and the preimages in (c)(d)
			depict the full orientation of the corresponding $\mathbf{m}$.
			The color sphere and the coordinate system are shown in the insets.}
		\label{hopfionnum}
	\end{figure}
	
	Figures 1(a) and 1(b) show the typical magnetization profiles of Bloch-type and N\'{e}el-type
	hopfions, respectively, obtained by numerical simulations.
	We consider a 16-nm-thick film with $A_{ex}=0.16$ pJ m$^{-1}$ and $M_s=1.51\times10^5$
	A m$^{-1}$, representing MnSi parameters \cite{magHp1}. No external field is applied.
	The Bloch-type (N\'{e}el-type) hopfions are favorable in bulk (interfacial) DMI systems.
	In Fig. 1(a), we use $K_s=0.5$ mJ m$^{-2}$, $K_b = 41$ kJ m$^{-3}$, and $D_b=0.115$ mJ m$^{-2}$,
	while in Fig. 1(b), we use $K_s=0.5$ mJ m$^{-2}$, $K_b = 20$ kJ m$^{-3}$, and $D_i=0.115$ mJ m$^{-2}$
	(these parameters are also used in the study of current-driven dynamics below).
	The simulations are mainly performed using mumax$^3$ \cite{mumax} at zero temperature
	(additional details of the simulations can be found in the Supplemental Materials \cite{SM}).
	We compute that the Hopf indices are 0.96 (Bloch) and 0.95 (N\'{e}el) by numerical integration of Eq. \eqref{hpindex}
	\cite{SM}. The two types of hopfions are topologically equivalent but behave
	differently in the presence of SHT, which we will discuss later.
	The upper and lower panels are the midplane cross-sections in the $xy$-plane
	and $xz$-plane. The magnetization profile in each $xy$-midplane cross-sections is
	Bloch-type (a) or N\'{e}el-type (b) skyrmionium or the target skyrmion \cite{TSK1,TSK2},
	while the $xz$-midplane cross-section shows a pair of vortices with opposite
	chirality. The right ($x>0$) $xz$-midplane contains a vortex (antivortex) with chirality $+1$ ($-1$)
	for an $H=+1$ ($H=-1$) hopfion.
	Outside the hopfions and at the center of the hopfions, the magnetization
	is along the $z$ direction, and the donut-shape transition region is chiral (for Bloch-type hopfions)
	or hedgehog-like (for N\'{e}el-type hopfions).
	Figure 1(c) and (d) show the corresponding preimages (constant-$\mathbf{m}$ curves in real
	space) of Fig. 1(a) and (b).
	The preimages link with each other once, which is consistent with the Hopf index calculation,
	justifying the hopfion nature of the textures in (a) and (b).
	
	Different from the hopfions observed in previous studies \cite{magHp1,magHp2,magHp3} that are
	confined in small magnetic disks, the introduction of a finite
	bulk PMA causes the hopfions in our work to be metastable, localized objects that can exist in
	long strips with a hopfion radius $R$, defined as the radius of the preimage $\mathbf{m}=(0,0,-1)$.
	Thus, these hopfions can be candidates of information carriers, and devices such as
	hopfion racetrack memories can be designed \cite{DW2, device1}. Moreover,
	unlike skyrmions, although the topology of a hopfion is nontrivial, the gyrovector
	$\mathbf{G}=\int \mathbf{F}\mathrm{d}V$ of a hopfion vanishes. Consequently, the main
	drawback of a FM skyrmion racetrack memory, the skyrmion Hall effect, is absent
	in the hopfion racetrack memory. In addition to the numerical verification, the vanishing gyrovector
	of a hopfion can be understood as follows. Consider a film that is isotropic in the $xy$ plane.
	The hopfion profile centered at a certain location can be expressed via $\Theta(r,\phi,z)$, $\Phi(r,\phi,z)$, where
	$(r,\phi,z)$ are cylindrical spatial coordinates, and $\Theta$, $\Phi$ are the polar
	and azimuthal angles of the magnetization. Because of the isotropy in the $xy$ plane,
	it is natural to assume that $\Theta$ is independent of $\phi$, and $\Phi(r,\phi,z)
	=\Delta\Phi(r,z)+n\phi$, where $n$ is an integer and $\Delta \Phi$ is a function
	independent of $\phi$. These assumptions are well justified by our numerical
	results. Thus, in cylindrical coordinates, $F_z=n\frac{\sin\Theta}{r}\frac{\mathrm{d}\Theta}{\mathrm{d} r}$.
	We can rewrite $G_z=\int F_z \mathrm{d}V $ as
	\begin{equation}
	G_z=\int_VF_zr\mathrm{d}r\mathrm{d}\phi \mathrm{d}z
	=-2n\pi \int_{-d/2}^{d/2} \left(\cos\Theta \big|^{r=\infty}_{r=0}\right)\mathrm{d}z.
	\end{equation}
	Since in a hopfion the magnetization directions are the same at both the periphery ($r=\infty$)
	and the center ($r=0$), $G_z$ vanishes. Since the two vortices in any $xz$ (or $yz$)
	midplane cross-section have opposite chirality, as shown in the lower panels of Fig. 1(a)(b),
	the integration of $F_x$ (or $F_y$) over the volume gives a vanishing contribution
	to $G_x$ (or $G_y$). The components of $\mathbf{G}$
	are invariant under continuous deformation \cite{SK2}; therefore, $\mathbf{G}=0$ applies to all the hopfions.
	
	\begin{figure}[!t]
		\begin{center}
			\includegraphics[width=8.5cm]{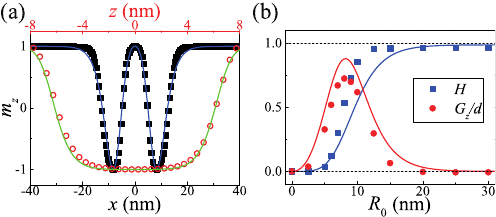}
		\end{center}
		\caption{
			(a) The profile of $m_z$ of the hopfion shown in Fig. 1(a). The bottom axis and
			black squares show the profile along the radial direction at $z=0$.
			The top axis and red circles show the profile along the $z$ direction at $r=R$.
			The solid lines are the ansatz \eqref{ansatz}.
			(b) The dependence of the Hopf index $H$ and layer-averaged gyrovector component $G_z/d$
			on the integration radius $R_0$. The symbols are numerical results, and the solid lines are
			obtained from the ansatz \eqref{ansatz}.}
		\label{profile}
	\end{figure}
	
	The magnetic hopfions discussed in previous studies \cite{magHp1,magHp2,magHp3}
	were Bloch-like. In the following, we mainly focus on N\'{e}el-type hopfions.
	Although the analytical expression of the hopfion profile is unknown, we find
	an ansatz that describes the $H=+1$ N\'{e}el-type hopfion profile very well:
	\begin{equation}\label{ansatz}
	\begin{gathered}
	m_x = \frac{4r^\prime \left[2z^\prime \sin\phi + \cos \phi \left({r^\prime}^2+{z^\prime}^2-1\right)\right]}{\left(1+{r^\prime}^2+{z^\prime}^2\right)^2}, \\
	m_y = \frac{4r^\prime \left[-2z^\prime \cos\phi + \sin \phi \left({r^\prime}^2+{z^\prime}^2-1\right)\right]}{\left(1+{r^\prime}^2+{z^\prime}^2\right)^2}, \\
	m_z = 1-\frac{8{r^\prime}^2}{\left(1+{r^\prime}^2+{z^\prime}^2\right)^2},
	\end{gathered}
	\end{equation}
	where $r^\prime=\frac{e^{R/w_R}-1}{e^{r/w_R}-1}$, $z'=
	\frac{z}{|z|}\frac{e^{|z|/w_h}-1}{e^{h/w_h}-1}$. $R$, $w_R$, $h$ and $w_h$ are lengths
	parametrizing the hopfion profile. $R$ is the hopfion radius, defined from $m_z(r=R,z=0)=-1$.
	$h$ is the hopfion height describing the extent of the hopfion in the out-of-plane direction,
	defined from $m_z(r=R, z=h)=1/9$.
	$w_R$ and $w_h$ are hopfion wall widths in the radial and out-of-plane directions
	respectively, describing the length scale of the magnetization
	variation from $m_z=+1$ to $m_z=-1$ \cite{SK5}.
	The ansatz \eqref{ansatz} is based on the well-known ansatz \cite{FHK}
	augmented by a non-linear rescaling of $r$ and $z$ \cite{SK5} and can also
	describe Bloch-type hopfions and $H=-1$ hopfions after simple transformations \cite{SM}.
	Figure 2(a) shows a comparison of $m_z$ between the above ansatz and the numerical data
	along the $x$ direction for $y=z=0$ (bottom axis) and along the $z$ direction for $r=R$ (top axis),
	with $R=8.3$ nm, $w_R=5.6$ nm, $h=6.3$ nm and $h_w=1.6$ nm obtained from fitting.
	The comparison gives good agreement (more comparisons can be found in the
	Supplemental Materials \cite{SM}). The numerical data along the $z$ direction are
	slightly asymmetric with respect to $z=0$, which is because of the asymmetric
	bulk magnetic charge. If the dipolar interaction is turned off, or if the hopfion
	is a Bloch-type hopfion, this asymmetry will vanish.
	
	Next, we numerically calculate the Hopf index $H$ and the layer-averaged gyrovector
	$G_z/d$ by integrating over a cylinder of height $d$ and radius $R_0$ (symbols), and we
	compare the numerical results with the analytical result calculated using the ansatz
	\eqref{ansatz} (solid lines), as shown in Fig. 2(b). As $R_0$ increases,
	$H$ converges toward 1, and $G_z/d$ converges toward 0. Note that the $R_0$
	used here is smaller than the sample size of our numerical simulation
	such that the edge structures are discarded.
	Below, we use this ansatz to discuss the current-driven dynamics of
	the hopfions, and we compare the results with numerical simulations.
	
	Disregarding deformations, the motion of a hopfion, as a rigid body, is governed by Thiele's
	equation \cite{Thiele,VT2}:
	\begin{equation}\label{thiele}
	\frac{\gamma}{M_s}\mathbf{T} + \mathbf{G}\times(\mathbf{v}-\mathbf{u})-\tensor{\mathcal{D}}\cdot(\alpha \mathbf{v}
	-\beta\mathbf{u}) = 0,
	\end{equation}
	where $\gamma$ is the gyromagnetic ratio;
	$\alpha$ is the Gilbert damping;  $\beta$ is the STT non-adiabaticity \cite{ZhangLi};
	$\mathbf{v}$ is the velocity of the hopfion;
	$\mathbf{u}=-\mu_{B}p\mathbf{J}/[eM_s(1+\beta^2)]$ is a vector with dimension of velocity proportional
	to the current density $\mathbf{J}$, in which $p$ is the spin polarization and
	$e$ is the electron charge; $\mathbf{G}$ is the above-mentioned gyrovector; and $\tensor{\mathcal{D}}$ is
	the dissipation tensor defined as $\mathcal{D}_{ij}=\int \partial_i\mathbf{m}\cdot\partial_j
	{\mathbf{m}}\mathrm{d}V$. $\mathbf{T}$ is the force on
	the hopfion, expressed as $T_i=-\frac{\partial \int \mathcal{F} \mathrm{d}V}{\partial X_i}-\int \frac{\partial \mathbf{m}}{\partial x_i}
	\cdot (\mathbf{m}\times\boldsymbol{\tau}) \mathrm{d}V$, where $\mathcal{F}$ is
	the free-energy functional \eqref{energy}, $X_i$ is the center position
	of the hopfion, and $ \boldsymbol{\tau}$ represents non-conservative torques other than STT
	such as the SHT. In our model, all the material parameters are
	spatially homogeneous; therefore, the first term in $\mathbf{T}$ is 0.
	Since $\mathbf{G}=0$, the hopfions move along the applied current via STT with velocity $\mathbf{v}=\frac{\beta}{\alpha}
	\mathbf{u}$. Figure 3(a) shows the trajectory during a period of 15 ns
	of the N\'{e}el-type hopfion driven by STT under $J=10^{11}$ A m$^{-2}$,
	with $p=0.12$ (a typical value for Co \cite{Copolar}), $\alpha=0.05$ and $\beta=0.1$,
	obtained by numerically solving the Landau-Lifshiz-Gilbert (LLG) equation \cite{LLG}
	with STT \cite{ZhangLi,mumax}.
	The strip is 128 nm-wide in the $y$ direction, and periodic
	boundary conditions are used in the $x$ direction.
	The trajectory is almost along the $x$ direction after moving for
	15 ns. The small deviation may come from the discretization and the deformation
	of the hopfion.
	Figure 3(b) shows the longitudinal component of the hopfion velocity $v_x$ versus the applied current density $J$.
	The numerical data from LLG simulations (black squares) are in good agreement with the analytical formula $v=\frac{\beta}{\alpha}u$
	(black line). Above $J=2\times 10^{11}$ A m$^{-2}$, the hopfion becomes
	distorted, and at even higher currents $J=5\times 10^{11}$ A m$^{-2}$, the hopfion is destroyed.
	In contrast to the threshold current for the annihilation of FM skyrmions,
	this limitation on the current is not intrinsic and can be improved by material engineering.
	For the Bloch-type hopfion in Fig. 1(a), similar results are obtained.
	
	\begin{figure}[!t]
		\begin{center}
			\includegraphics[width=8.5cm]{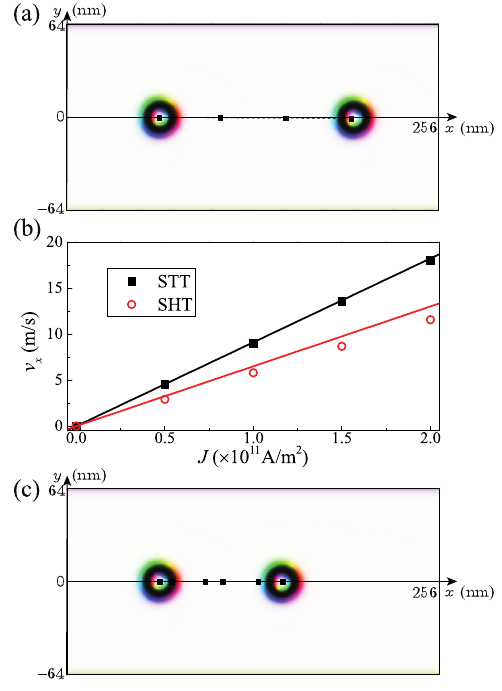}
		\end{center}
		\caption{
			(a) Trajectory of N\'{e}el hopfion driven by STT during a period of 15 ns.
			The midplane cross-section of $\mathbf{m}$ in the $xy$-plane is shown.
			(b) Current density $J$ dependence of the longitudinal velocity $v_x$ of the N\'{e}el hopfion.
			The black squares (red circles) are numerical results for STT-driven (SHT-driven) motion. The solid lines
			are theoretical predictions.
			(c) Trajectory of N\'{e}el hopfion driven by SHT during 15 ns.
			The midplane cross-section of $\mathbf{m}$ in the $xy$ plane is shown.
			The color map of (a) and (c) is the same as in Fig. 1.}
		\label{dyn}
	\end{figure}
	
	Recently, spin-orbit torques (SOTs) have attracted attention for
	driving magnetic textures because of their possibly higher angular momentum transfer
	efficiency \cite{SOTReview}. SOTs arise from a variety of origins
	such as interfacial Rashba spin-orbit coupling  \cite{Manchon},
	spin-Hall-effect-induced spin currents from adjacent heavy metal
	layers \cite{SHT}, and the intrinsic SOT in  magnetic materials \cite{Hals}.
	The field-like component of the torque \cite{Manchon,Hals} can be regarded as a uniform magnetic field
	on the system. Since a hopfion is a localized object
	in a domain, a uniform magnetic field deforms (or even destroys) the hopfion
	without exerting a net force on it. We consider the antidamping-like SHT \cite{SHT,ZY},
	\begin{equation} \label{spinhall}
	\boldsymbol{\tau}=\frac{\gamma \hbar}{eM_sd}\theta_\text{SH} \mathbf{m}\times[\mathbf{m}\times
	(\hat{\mathbf{J}}\times \hat{\mathbf{z}})],
	\end{equation}
	which is usually the dominant SOT for a heavy metal/magnet system
	where $\theta_\text{SH}$ is the spin Hall angle. Consider a current applied along the $x$ direction.
	The SHT is then $\boldsymbol{\tau}=\tau_0 \mathbf{m}\times(\mathbf{m}\times \hat{\mathbf{y}})$,
	where $\tau_0$ denotes the prefactors in \eqref{spinhall}.
	Using the ansatz \eqref{ansatz} with $R$, $w_R$, $h$ and $w_h$ obtained
	by fitting the numerical data, we can calculate the force $\mathbf{T}$ and dissipation
	tensor $\tensor{\mathcal{D}}$. According to the polarity of the hopfion profile,
	the force on a Bloch-type hopfion is along the $y$ direction, while the force
	on a N\'{e}el-type hopfion is along the $x$ direction, similar to the skyrmion
	or target skyrmion \cite{Hall2,TSK1}. Thus, only N\'{e}el
	hopfions move along the current under SHT, while the Bloch hopfions
	move transverse to the current and are blocked by the edge of the racetrack.
	Because of the isotropy in the $xy$ plane, $\tensor{\mathcal{D}}$ is
	diagonal, with $\mathcal{D}_{xx}=\mathcal{D}_{yy}\equiv\mathcal{D}$.
	Thus, we have $v_x=\frac{T}{\alpha \mathcal{D}}$ for N\'{e}el hopfions.
	The trajectory of the N\'{e}el-type hopfion during 15 ns
	driven by SHT under $J=10^{11}$ A m$^{-2}$ and $\theta_\text{SH}=0.05$
	(a typical value for Pt \cite{SHT}) obtained from the LLG simulation is shown in Fig. 3(c). The damping is
	assumed to be $\alpha=0.05$. The N\'{e}el hopfion propagates along the wire.
	The longitudinal velocity component $v_x$ under different current densities
	is plotted in Fig. 3(b) by red circles. The analytical formula (red line) agrees
	well with the numerical data. Note that the values of $T$ and $\mathcal{D}$ depend
	on the hopfion profile. Since the ansatz introduced gives very good agreement
	with the numerical results, it may be useful in other investigations
	on hopfions.
	
	Note that hopfions can also be stabilized in AFM systems,
	where the staggered N\'{e}el field
	forms a hopfion profile \cite{SM,unpub}.
	
	The N\'{e}el-type hopfions should be realizable in experiments \cite{SM,multilayer1,multilayer2,multilayer3,CoAni}.
	In device application, a hopfion can be created by applying a
	spin-polarized current or a localized magnetic field through a ring-shaped nanocontact
	\cite{SM,TSK1,SK3,SKAQ1,SKAQ2,heat}. A strong out-of-plane magnetic field
	can eliminate a hopfion. The creation and elimination of hopfions will be studied in
	detail in future. Since the hopfions have finite magnetic moment, any existing
	techniques that can detect local magnetic moment are also capable to detect
	hopfions \cite{magHp1,elec,NV}.
	The hopfions that we found remain geometrically confined by the thickness of the film
	with the help of strong PMA.
	Indeed, in the presence of DMI,  Derrick's theorem \cite{Derrick},
	which prohibits the existence of 3D solitons in infinite conventional
	(non-chiral) magnets, is no longer valid \cite{Bog1995}. Whether it is possible to stabilize
	hopfions in 3D chiral magnets without confinement is still
	an open question for further investigations.
	Our study also implies that magnetic
	systems represent a fertile playground for research on nonlinear 3D topological solitons.
	
	In conclusion, we identified a new type of hopfion, the N\'{e}el-type hopfion, and studied the current-driven dynamics of hopfions. In FM systems, despite the nontrivial topology, neither Bloch- nor N\'{e}el-type hopfions exhibit Hall effects and propagate along external currents via spin transfer torque. The SHT only drives the N\'{e}el-type hopfions to move along the current. Hopfions have the potential to be efficient information carriers.
	
	\begin{acknowledgments}
		The research leading to these results was supported by the European Research Council via Advanced Grant No. 669442, ``Insulatronics,''
		and by the Research Council of Norway through its Centres of Excellence funding scheme, Project No. 262633, ``QuSpin.''
		X.S.W. acknowledges the
		support from the Natural Science Foundation of China (Grant No. 11804045)
		and the China Postdoctoral Science Foundation (Grant No. 2017M612932 and 2018T110957).
	\end{acknowledgments}

\clearpage
\onecolumngrid
\section{Supplemental Materials}
\setcounter{table}{0}
\setcounter{equation}{0}
\renewcommand\theequation{S\arabic{equation}}
\renewcommand*{\citenumfont}[1]{S#1}
\renewcommand*{\bibnumfmt}[1]{S#1}
\setcounter{figure}{0}
\subsection{Simulation Details}
Most of the simulations are performed using the mumax$^3$ package \cite{mumax1}. Some of the
results of the static hopfion profile are double-checked by the \textsc{oommf} package \cite{oommf}.
All the calculates are performed at zero temperature. The mesh size is 0.5 nm$\times$0.5 nm$\times$0.5 nm.
The surface pinning is modeled by
imposing a very strong PMA $K=10^6$ J m$^{-3}$ on two additional layers attached to the top
and bottom surfaces. This corresponds to a surface anisotropy $K_s=0.5$ mJ m$^{-2}$ by
multiplying the mesh size.

For the static hopfion profile, the conjugate gradient method \cite{CG} is used to minimize the total
energy with an error toleration of $10^{-5}$. The sample size is 128 nm$\times$128 nm$\times$16 nm,
as shown in Fig. S1. To be consistent with the current-driven dynamical
simulations, periodical boundary conditions are imposed along the $x$ direction to mimic
a long strip along the $x$ direction. Two sets of initial magnetizations are
used. One magnetization is a ring of $\mathbf{m}=(0,0,-1)$ at 25 nm$\leq r\leq$ 40 nm and $|z|<5$ nm inside
a uniform domain of $\mathbf{m}=(0,0,1)$. The other magnetization is a profile of a well-known ansatz
that will be discussed below. Both sets of initial magnetizations give the same results.
For the current-driven dynamics,
the RK45 method is used for the temporal integration of the Landau-Lifshitz-Gilbert (LLG) equation \cite{LLG}.
The spin-transfer torque is in the Zhang-Li form \cite{ZhangLi2,mumax1}.

Exchange constant $A_{ex}=0.16$ pJ m$^{-1}$ and saturation magnetization $M_s=1.51\times10^5$ A m$^{-1}$,
are used throughout the paper, representing MnSi parameters.
Other material parameters used are $K_s=0.5$ mJ m$^{-2}$, $K_b = 41$ kJ m$^{-3}$ and $D_b=0.115$ mJ m$^{-2}$
[for the Bloch-type hopfion shown in Fig. 1(a)], and
$K_s=0.5$ mJ m$^{-2}$, $K_b = 20$ kJ m$^{-3}$, and $D_i=0.115$ mJ m$^{-2}$
[for the N\'{e}el-type hopfion shown in Fig. 1(b) as well as Fig. 2 and Fig. 3.]

\subsection{Calculation of Hopf Index}
\noindent\textbf{Analytical discussions-}As mentioned in the main text, for an infinite system, the Hopf index
is defined as
\begin{equation}
H=\frac{1}{(4\pi)^2}\int\mathbf{F}\cdot\mathbf{A}\mathrm{d}V,
\end{equation}
where $F_i=\frac{1}{2}\varepsilon_{ijk}\mathbf{m}\cdot\left(\partial_j \mathbf{m}\times\partial_k \mathbf{m}\right)$, in which $i,j,k=\{x,y,z\}$ and
$\varepsilon$ is the Levi-Civita tensor, and $\mathbf{A}$
is a vector potential satisfying $\nabla\times \mathbf{A}=\mathbf{F}$.

We now demonstrate that the Hopf index is well-defined for an infinite system.
Straightforward derivation shows that $\mathbf{F}$ is divergenceless ($\nabla\cdot\mathbf{F}=0$)
when $|\mathbf{m}|=$constant such that the vector potential $\mathbf{A}$ exists.
However, obviously, $\mathbf{A}$ is not unique. For any continuous function
$\varphi(\mathbf{r})$, $\mathbf{A}^\prime=\mathbf{A}+\nabla \varphi$ is also a vector potential.
The corresponding Hopf index is
\begin{equation}
H^\prime=\frac{1}{(4\pi)^2}\int\mathbf{F}\cdot\mathbf{A}^\prime \mathrm{d}V=H+\frac{1}{(4\pi)^2}\int\mathbf{F}\cdot\nabla\varphi \mathrm{d}V.
\end{equation}
The integral in the extra term can be rewritten as
\begin{equation}
\int\mathbf{F}\cdot\nabla\varphi \mathrm{d}V=\int \nabla\cdot(\varphi\mathbf{F}) \mathrm{d}V -\int \varphi\nabla\cdot\mathbf{F} \mathrm{d}V
=\oint \varphi\mathbf{F} \cdot d\mathbf{S} -0=\oint \varphi\mathbf{F} \cdot d\mathbf{S},
\end{equation}
where Gauss's theorem has been used and $\oint$ means the integration over the
surface of the volume. In an infinite system, the surface is infinitely far
away, and the $\mathbf{m}$ field should be homogenous such that, on the surface,
$\mathbf{F}$ is 0, and the integral $\oint \varphi\mathbf{F} \cdot d\mathbf{S}$ vanishes. Thus, we have
$H^\prime=H$, meaning that the Hopf index is well-defined independent
of the choice of $\mathbf{A}$.

For a rotationally symmetric system, it is natural
to assume that the hopfion profile following $\Theta$ is independent of $\phi$, and $\Phi(r,\phi,z)
=\Delta\Phi(r,z)+n\phi$, where $n$ is an integer. This form means that
the polar angle (or $z$ component) of $\mathbf{m}$ is independent of $\phi$, and
when transversing a whole circle centered at the origin in real space ($\phi$ changes from 0
to $2\pi$), the azimuthal angle $\Phi$ of $\mathbf{m}$ uniformly rotates by $2n\pi$.
With this assumption, we can  write the $\mathbf{F}$ field in cylindrical
coordinates as
\begin{gather}
	F_r=\mathbf{m}\cdot\left(\frac{\partial\mathbf{m}}{r\partial \phi}\times\frac{\partial\mathbf{m}}{\partial z}\right)=-n\frac{\sin\Theta}{r}\frac{\partial \Theta}{\partial z},\\
	F_\phi=\mathbf{m}\cdot\left(\frac{\partial\mathbf{m}}{\partial z}\times\frac{\partial\mathbf{m}}{\partial r}\right)
	=\sin\Theta \left(\frac{\partial\Theta}{\partial z}\frac{\partial\Delta\Phi}{\partial r}-\frac{\partial\Theta}{\partial r}\frac{\partial\Delta\Phi}{\partial z}\right),\\
	F_z=\mathbf{m}\cdot\left(\frac{\partial\mathbf{m}}{\partial z}\times\frac{\partial\mathbf{m}}{r\partial \phi}\right)=n\frac{\sin\Theta}{r}\frac{\partial \Theta}{\partial r}.
\end{gather}
The vector potential $\mathbf{A}$ is
\begin{gather}
	A_r=-(1+\cos\Theta)\frac{\partial\Delta\Phi}{\partial r},\\
	A_\phi=\frac{n}{r}(1-\cos\Theta),\\
	A_z=-(1+\cos\Theta)\frac{\partial\Delta\Phi}{\partial z}.
\end{gather}
Then, the Hopf index is
\begin{equation}
H=\frac{1}{(4\pi)^2}\int \mathbf{F}\cdot\mathbf{A}\mathrm{d}V=\frac{n}{4\pi}\int_{-\infty}^{+\infty}
\int_0^{+\infty}\sin\Theta\left(\frac{\partial\Theta}{\partial z}\frac{\partial\Delta\Phi}{\partial r}
-\frac{\partial\Theta}{\partial r}\frac{\partial\Delta\Phi}{\partial z}\right)\mathrm{d}r\mathrm{d}z.
\label{cylindhopf}
\end{equation}
Thus, the Hopf index equals  the whirling number $n$ along the $\phi$ direction
multiplied by the skyrmion number at the $rz$ half plane \cite{HPindex}.

\begin{figure}[hbp]
	\renewcommand\thefigure{S\arabic{figure}}
	\centering
	\includegraphics[width=12.5cm]{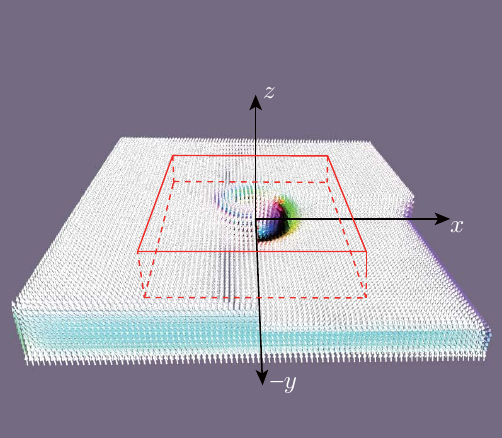}\\
	\caption{A sample of the simulation for the static hopfion. One octant is made transparent
		to visualize the magnetization profile inside. The volume in the Hopf index
		calculation is indicated by the red box. }
	\label{FigS1}
\end{figure}

\noindent\textbf{Numerical evaluation of Hopf index-} As discussed above, the Hopf index
is well-defined when $\oint \left(\varphi\mathbf{F}\right) \cdot d\mathbf{S}=0$ is satisfied.
To numerically evaluate the Hopf index,
we first cut off the nonhomogeneous edge such that $\mathbf{m}$ is homogeneous
on the surface of the sample to ensure that the Hopf index is well-defined
(see Fig. S1). We then employ two methods to calculate the Hopf index: a real space
method and a Fourier space method.

In real space, we first numerically calculate $\mathbf{F}$ from $\mathbf{m}$
utilizing the standard central finite difference method. Then, we employ a
radial basis function (RBF) interpolation to ensure that $\mathbf{F}$
is divergenceless \cite{RBF}. We use a Gaussian function $g(\mathbf{x}_i,\mathbf{x}_j)
=e^{-\epsilon \left(|\mathbf{x}_i-\mathbf{x}_j|^2\right)}$ as the RBF
with control parameter $\epsilon=1$. Here, $i$ and $j$ label two grid points, and $\mathbf{x}_i$ and   $\mathbf{x}_j$
denote the positions of $i$ and $j$. The interpolated field is
\begin{equation}
\mathbf{F}(\mathbf{x})=\sum_i \left(\nabla\nabla-\nabla^2\right)g\left(\mathbf{x},\mathbf{x}_i\right)\mathbf{c}_i.
\end{equation}
After obtaining the RBF coefficients $\mathbf{c}_i$, $\mathbf{A}$ can be directly
calculated:
\begin{equation}
\mathbf{A}(\mathbf{x})=-\sum_i (\mathbf{c}_i\times\nabla)g\left(\mathbf{x},\mathbf{x}_i\right),
\end{equation}
where the Coulomb gauge is used. Then, the standard numerical integration is performed to calculate
$H=\frac{1}{\left(4\pi\right)^2}\int \mathbf{F}\cdot\mathbf{A} \mathrm{d}V$.

The Fourier space method has been introduced in Ref. \cite{magHp22}. The relative difference
between the two methods is less than 5\%. In the main text, we
show the Fourier space result.

\subsection{Construction and Verification of the Ansatz for Hopfion Profile}
We start from the well-known ansatz for hopfions \cite{ansatz12}:
\begin{gather}
	m_x=\frac{4\left[2xz-y\left(x^2+y^2+z^2-1\right)\right]}{\left(1+x^2+y^2+z^2\right)^2},\\
	m_y=\frac{4\left[2yz+x\left(x^2+y^2+z^2-1\right)\right]}{\left(1+x^2+y^2+z^2\right)^2},\\
	m_z=1-\frac{8\left(x^2+y^2\right)}{\left(1+x^2+y^2+z^2\right)^2},
\end{gather}
which describes an $H=-1$ Bloch-type hopfion with $\mathbf{m}$ upward at the center and infinity.
Note that in some references, the definitions of Hopf index differ by a sign. Here, we use the
definition mentioned in the main text.
The radius is 1 because at $r^2=x^2+y^2=1$, $\mathbf{m}=(0,0,-1)$.
The magnetization rotates counterclockwise (clockwise) at $r>1$ ($r<1$) in the top view.
An $H=+1$ hopfion ansatz can be obtained simply by inverting the sign of
the first term in the numerators of $m_x$ and $m_y$:
\begin{gather}
	m_x=\frac{4\left[-2xz-y\left(x^2+y^2+z^2-1\right)\right]}{\left(1+x^2+y^2+z^2\right)^2},\\
	m_y=\frac{4\left[-2yz+x\left(x^2+y^2+z^2-1\right)\right]}{\left(1+x^2+y^2+z^2\right)^2},\\
	m_z=1-\frac{8\left(x^2+y^2\right)}{\left(1+x^2+y^2+z^2\right)^2}.
\end{gather}
In cylindrical coordinates:
\begin{gather}
	m_x=\frac{4r\left[-2\cos\phi z-\sin \phi \left(r^2+z^2-1\right)\right]}{\left(1+r^2+z^2\right)^2},\\
	m_y=\frac{4r\left[-2\sin\phi z+\cos \phi \left(r^2+z^2-1\right)\right]}{\left(1+r^2+z^2\right)^2},\\
	m_z=1-\frac{8r^2}{\left(1+r^2+z^2\right)^2}.
\end{gather}
This operation can invert the
Hopf index and retain the rotational sense, which is preferred for $D_b>0$. For $D_b<0$,
we let $m_x\rightarrow-m_x$ and $m_y\rightarrow-m_y$. For a realistic hopfion
with a given radius $R$ and height, it is natural to consider a linear rescaling
, where $h$ describes the extent
of the hopfion in the $z$ direction. This profile is used as the initial condition,
with $R=20$ nm and $h=10$ nm. However, this ansatz cannot describe the numerical
data well. In Fig. S2(a), we show the numerical magnetization profile along $x$ at $y=z=0$ (symbols),
and the above ansatz (dashed blue line) with $R=7.8$ nm obtained from the numerical data for the Bloch-type
hopfion is shown in Fig. 1(a) of the main text.
Obviously, the numerical data show a much faster decay away from $x=R$ than the
polynomial decay of the ansatz. Inspired by  work on skyrmion profiles \cite{wxs2},
we introduce another length scale to describe how fast $\mathbf{m}$ decays to $(0,0,1)$.
Since the $z$ direction is special because of the surface PMA, we introduce $w_R$ and $w_h$
for the $xy$ plane and $z$ direction, respectively. We try a monotonic nonlinear rescaling
\begin{equation}
r\rightarrow r^\prime=\frac{e^{r/w_R}-1}{e^{R/w_R}-1},\quad z\rightarrow z^\prime=\frac{|z|}{z}\frac{e^{|z|/w_h}-1}{e^{h/w_h}-1},
\end{equation}
and use the resulting ansatz to fit the numerical result to determine the parameters $R$, $w_R$,
$h$, and $w_h$. $R$ and $w_R$ are determined by fitting the radial profile $m_z(r)$ at $z=0$. $h$ and $w_h$ are
determined by fitting the profile $m_z(z)$ along $z$ at $r=R$. The result of this nonlinear rescaling
(solid red line) is also compared with the numerical data in Fig. S2(a). The agreement is obversely better.
For a larger hopfion (which can be obtained by using a smaller $K_b$), the agreement of our ansatz is even better.
\begin{figure}[hbp]
	\renewcommand\thefigure{S\arabic{figure}}
	\centering
	\includegraphics[width=18cm]{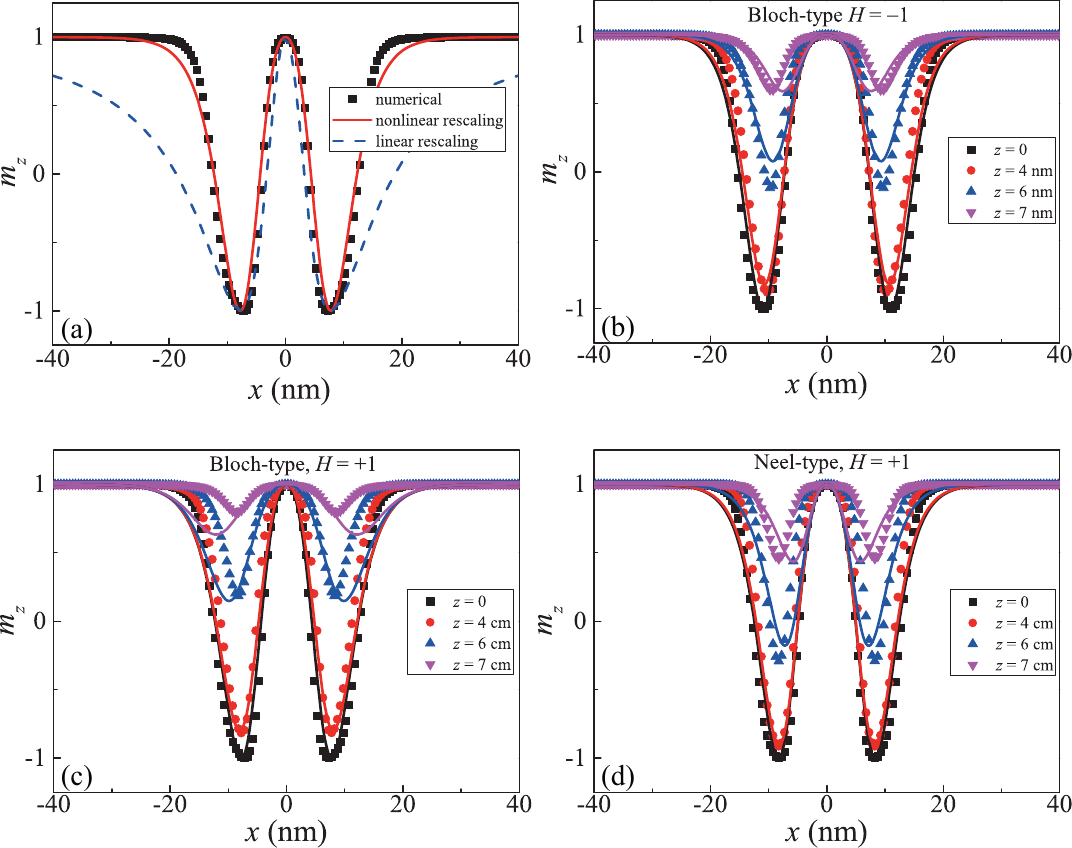}\\
	\caption{(a) Comparison of $m_z(x)$ at $y=0$, $z=0$ between numerical data (symbols) and
		ansatz (lines). The solid red line is the result
		of the nonlinear rescaling shown here. The dashed blue line is the linear rescaling.
		(b)(c)(d) Comparison of $m_z(x)$ at $y=0$ for different $z$. The symbols
		are numerical data, and the solid lines of the same color are the results of the ansatz.
		(b) $H=+1$ Bloch-type hopfion ($K_b=41$ kJ m$^{-3}$, $D_b=0.115$ mJ m$^{-2}$).
		(c) $H=-1$ Bloch-type hopfion ($K_b=39$ kJ m$^{-3}$, $D_b=0.115$ mJ m$^{-2}$).
		(d) $H=+1$ N\'{e}el-type hopfion ($K_b=20$ kJ m$^{-3}$, $D_i=0.115$ mJ m$^{-2}$).}
	\label{FigS2}
\end{figure}

\begin{figure}[hbp]
	\renewcommand\thefigure{S\arabic{figure}}
	\centering
	\includegraphics[width=18cm]{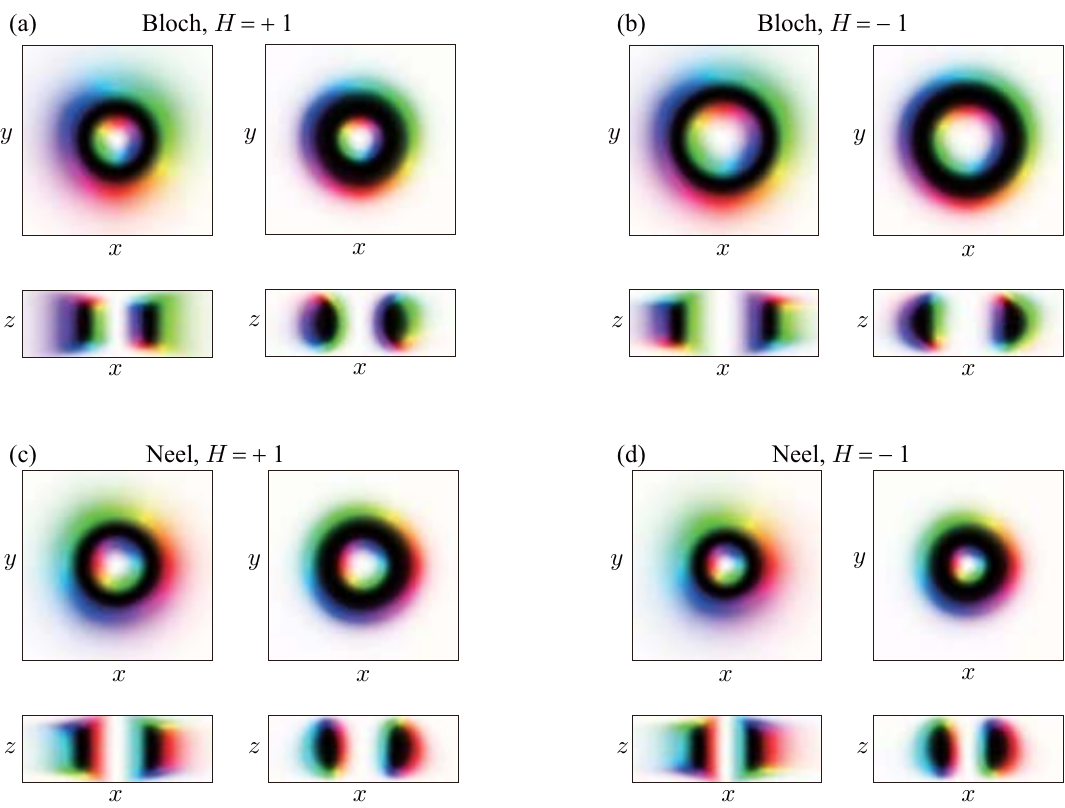}\\
	\caption{Comparison of $\mathbf{m}$ between numerical data and
		ansatz for (a) $H=+1$ Bloch-type hopfion ($K_b=41$ kJ m$^{-3}$, $D_b=0.115$ mJ m$^{-2}$),
		(b) $H=-1$ Bloch-type hopfion ($K_b=39$ kJ m$^{-3}$, $D_b=0.115$ mJ m$^{-2}$),
		(c) $H=+1$ N\'{e}el-type hopfion ($K_b=20$ kJ m$^{-3}$, $D_i=0.115$ mJ m$^{-2}$),
		(d) $H=-1$ N\'{e}el-type hopfion ($K_b=20$ kJ m$^{-3}$, $D_i=0.115$ mJ m$^{-2}$).}
	\label{FigS3}
\end{figure}

For an $H=+1$ hopfion, with $z\neq0$, the position of the minima of $m_z(r)$ moves outward as $|z|$ increases
according to the ansatz, which is consistent with the numerical result, as shown in Fig. S2(b).
In contrast, for an $H=-1$ hopfion, the minima of $m_z(r)$ moves inward, as shown in Fig. S2(c).
To describe this, we further invert the rescaling of $r$ as
\begin{equation}
r\rightarrow r^\prime=\left(\frac{e^{r/w_R}-1}{e^{R/w_R}-1}\right)^{-1}=\frac{e^{R/w_R}-1}{e^{r/w_R}-1},
\end{equation}
which maps $r=0$ to $r^\prime=\infty$ and vice versa.
In cylindrical coordinates for space and spherical coordinates for $\mathbf{m}$,
the ansatz can be written as
\begin{gather}
	\cos\Theta = 1-\frac{8{r^{\prime}}^2}{\left(1+{r^{\prime}}^2+{z^{\prime}}^2\right)^2},\\
	\Phi=\phi+\arctan \left(-\frac{1+{r^{\prime}}^2+{z^{\prime}}^2}{2z^\prime}\right).
\end{gather}
The Hopf index can be calculated using Eq. \eqref{cylindhopf}.
Because of the inverse rescaling of $r$, the
Hopf index becomes $1$. To retrieve the rotational sense, we further
let $m_x\rightarrow -m_x$, $m_y\rightarrow -m_y$, or $\Phi\rightarrow \Phi+\pi$.
The resultant ansatz gives good agreement for
$z\neq0$, as shown in Fig. S2(b) by the solid lines. Although the ansatz cannot quantitatively
overlap with the numerical data, the tendency of the minima position is correct.

To verify the continuity of the ansatz, we write it in Cartesian coordinates as
\begin{gather}
	m_x = \frac{4r^\prime \left[-2z^\prime \frac{x}{\sqrt{x^2+y^2}} - \frac{y}{\sqrt{x^2+y^2}}  \left({r^\prime}^2+{z^\prime}^2-1\right)\right]}{\left(1+{r^\prime}^2+{z^\prime}^2\right)^2}, \\
	m_y = \frac{4r^\prime \left[-2z^\prime \frac{y}{\sqrt{x^2+y^2}} + \frac{x}{\sqrt{x^2+y^2}} \left({r^\prime}^2+{z^\prime}^2-1\right)\right]}{\left(1+{r^\prime}^2+{z^\prime}^2\right)^2}, \\
	m_z = 1-\frac{8{r^\prime}^2}{\left(1+{r^\prime}^2+{z^\prime}^2\right)^2},
\end{gather}
where  $r^\prime=\left(\frac{e^{\sqrt{x^2+y^2}/w_R}-1}{e^{R/w_R}-1}\right)^{\pm1}$.
When $xyz\neq0$, the ansatz has no singularity. At $z=0$, we have
$\lim_{z\rightarrow 0}z^\prime =0$ and $\partial_z z^\prime\big|_{0+}=\partial_z z^\prime\big|_{0-}$.
At $x=y=z=0$, it is easy to verify $\lim_{x,y,z\rightarrow 0}\mathbf{m}=(0,0,1)$,
and all the first-order derivatives $\partial_im_j$ ($i,j\in x,y,z$) are
continuous. Since the highest order of derivative in the energy functional
is 1, the ansatz is well-defined in the whole space.

To obtain a N\'{e}el-type hopfion, we can locally rotate the $\mathbf{m}$ by 90
degrees, as shown in the main text:
\begin{equation}
\begin{gathered}
m_x = \frac{4r^\prime \left[2z^\prime \sin\phi + \cos \phi \left({r^\prime}^2+{z^\prime}^2-1\right)\right]}{\left(1+{r^\prime}^2+{z^\prime}^2\right)^2}, \\
m_y = \frac{4r^\prime \left[-2z^\prime \cos\phi + \sin \phi \left({r^\prime}^2+{z^\prime}^2-1\right)\right]}{\left(1+{r^\prime}^2+{z^\prime}^2\right)^2}, \\
m_z = 1-\frac{8{r^\prime}^2}{\left(1+{r^\prime}^2+{z^\prime}^2\right)^2}.
\end{gathered}
\end{equation}
Because the interfacial DMI only exists in the $xy$ plane and because there is no chiral interaction
in the $z$ direction, the minima in $m_z(r)$ remains in the same position. Therefore, either
$r^\prime = frac{e^{R/w_R}-1}{e^{r/w_R}-1}$ or $r^\prime=\frac{e^{r/w_R}-1}{e^{R/w_R}-1}$ can be used
as the ansatz. We choose $r^\prime = frac{e^{R/w_R}-1}{e^{r/w_R}-1}$ because
it gives better results in the calculation of $\tensor{D}$ and $\mathbf{T}$ [Fig. S2(d)].

The midplane cross-sections in the $xy$ plane and $yz$ plane are compared in Fig. S3(a-d)
for $H=+1$ Bloch-type, $H=-1$ Bloch-type, $H=+1$ N\'{e}el-type, and $H=-1$ N\'{e}el-type hopfions.
The 2D cross-sections also give fairly good agreement with the numerical results.
The senses of the rotation of the spins are retrieved in all the cross-sections, while the shapes of
the textures in $yz$ cross-sections are not as good as those in the $xy$ cross-sections.
Nevertheless, the ansatz can well describe the topological properties and current-driven
dynamics of the hopfions.

\subsection{Antiferromagnetic Hopfions}
\begin{figure}[hbp]
	\renewcommand\thefigure{S\arabic{figure}}
	\centering
	\includegraphics[width=15cm]{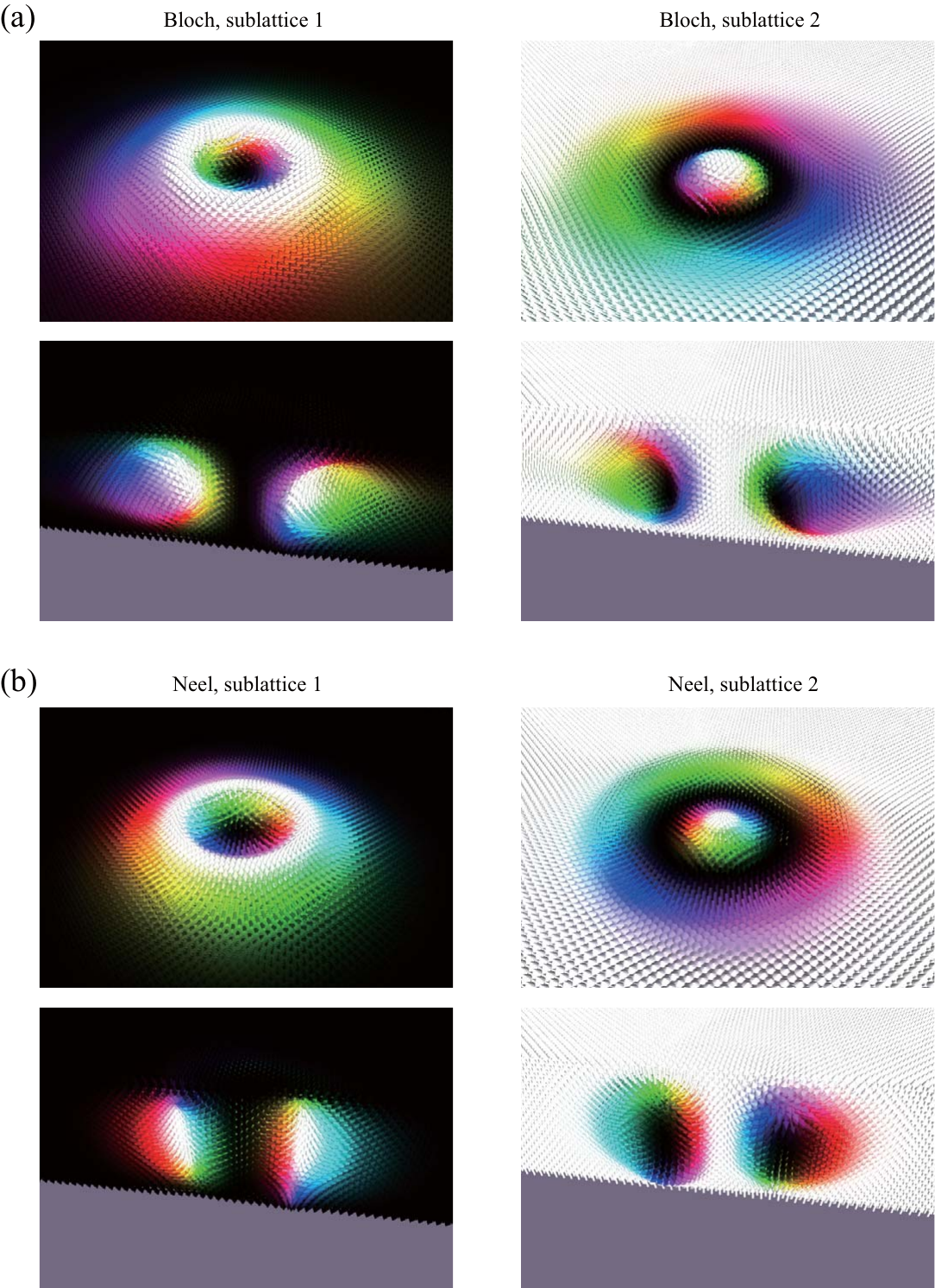}\\
	\caption{Mid-plane cross-sections in the $xy$ plane (upper) and $xz$ plane (lower) of sublattices 1 (left) and 2 (right).
		(a) Bloch-type AFM hopfion. (b) N\'{e}el-type AFM hopfion.}
	\label{FigS4}
\end{figure}

We consider that $A_{ex}=-0.16$ pJ m$^{-1}$, $D_b=0.115$ mJ m$^{-2}$, $K_b=16$ kJ m$^{-3}$,
and the other parameters to be the same as those in Fig. 1(a).
The ground state is an out-of-plane AFM N\'{e}el state. Because the dipolar field
is negligible in an antiferromagnet, to speed up the simulation, we
turn off the dipole-dipole interaction. For a numerical cell labelled by $(i,j,k)$,
if $i+j+l$ is even (sublattice 1), we impose the ansatz [Eq. (S19-S22)] with the
collective coordinates  $R = 15$ nm, $w_R = 5$ nm, $h = 8$
nm and $h_w = 5$ nm. If $i+j+k$ is odd (sublattice 2), we impose the opposite direction.
After relaxation, we obtain an antiferromagnetic Bloch-type hopfion.
In Fig. \ref{FigS4}(a), the mid-plane cross-sections in the $xy$ and $xz$ planes
are shown for each sublattice.
If we use $K_b=5$ kJ m$^{-3}$ and $D_i=-0.115$ mJ m$^{-2}$ instead, a N\'{e}el-type
AFM hopfion is obtained, as shown in Fig. \ref{FigS4}(b).

\subsection{Discussions about Experimental Realization, Writing, and Reading of Magnetic Hopfions}

\textit{Candidate materials-} The possible material for realization of Bloch-type
hopfions has been discussed in Ref.\cite{maghopf1}. For N\'{e}el-type hopfions
proposed in our paper, inversion-symmetry-breaking multilayer systems \cite{multilayer1,multilayer2}
may be good candidates for experimental realization.
Inversion-symmetry-breaking multilayer stacks such as Pt/Co/AlO$_x$
has been shown to possess interfacial DMI and support magnetic skyrmions.
The interlayer coupling is RYYK-type and can be tuned to be ferromagnetic.
The anisotropy of the ferromagnetic layers  (usually Co or Co alloys)
can be engineered in the wide range, from negative PMA (in-plane anisotropy)
to very strong PMA \cite{CoAni}. Thus, the weak bulk PMA and strong surface PMA
are possible in these systems. In this case, the spin Hall torque is much larger
than that calculated in the main text because
the torque is exerted on layers with substantially smaller thickness $d$.
Thus, the expected hopfion velocity can be much larger.

\textit{Writing of hopfion bits-} To use hopfions as information carriers,
we need to create or eliminate a hopfion to write binary ``1" or ``0".
We expect that Hopfions can be created from a single
domain by reversing the spins in a torus in analogous ways as for skyrmions.
The reversal can be achieved by applying a field pulse \cite{field1,field2}
or a current pulse \cite{Roland,TSK1}, and may be assisted by local heating \cite{heat}.
The field or current pulse can be applied through a ring-shaped nanocontact with
radius closed to the hopfion radius, with the field direction or spin-polarization
direction opposite to the magnetization direction of the domain. Since the top
and bottom surfaces are pinned by the strong PMA while the bulk PMA is much weaker,
a moderate field or current pulse can reverse the magnetization only in a torus.
After relaxation, the structure is able to evolve to a hopfion state. A hopfion
can be eliminated simply by applying a uniform magnetic field that is strong
enough to overcome the energy barrier between the hopfion state and the single-domain state.
The optimal pulse duration and field/current intensity for hopfion
creation depend strongly on the material parameters, and extensive numerical calculations are
needed. We intend to study the writing and erasing of hopfions in details elsewhere.

\textit{Reading of hopfion bits-} Although hopfions are local solitons in a uniform domain,
they have non-zero net magnetic moments. Thus, any existing techniques that can detect
local magnetic moments can be used to read the hopfion information. For example,
the possible Lorentz TEM image has been shown in Ref. \cite{maghopf1}. Other techniques
like magnetic force microscopy (MFM) also works for reading of hopfions.
More practically, the all-electric detection \cite{elec}
based on the non-collinear magnetoresistance effects is also possible. The use of
NV-centers should be able to detect the stray fields from the Hopfions \cite{NV}.

\end{document}